\documentclass[12pt,a4paper]{article}
\usepackage{amsfonts}
\usepackage{amssymb}
\usepackage{amsmath}
\usepackage{graphicx,color}
\usepackage{amsthm}
\usepackage[latin9]{inputenc}
\usepackage{enumerate}

\numberwithin{equation}{section}

\date{}
\begin{document}

\title{Double-delta potentials: one dimensional scattering. The Casimir effect and kink fluctuations.}

\author{J. Mateos Guilarte$^{1,2}$\footnote{guilarte@usal.es}, and J. M. Mu$\tilde{\rm n}$oz-Casta$\tilde{\rm n}$eda$^3$\footnote{jose.munoz-castaneda@uni-leipzig.de}\\
\footnotesize{{\sl $^1$Departamento de F\'{\i}sica Fundamental, Universidad de Salamanca, Spain.}}\\
\footnotesize{{\sl $^2$IUFFyM, Universidad de Salamanca, Spain.}}\\
\footnotesize{{\sl $^3$Institut f\"ur Theoretische Physik, Universit\"at Leipzig, Germany.}}}
\maketitle

\maketitle

\begin{abstract}
The path is explored between one-dimensional scattering through Dirac-$\delta$ walls and one-dimensional quantum field theories defined on a finite length interval with Dirichlet boundary conditions. It is found that two $\delta$'s are related to the Casimir effect whereas two $\delta$'s plus the first transparent P$\ddot{\rm o}$sch-Teller well arise in the context of the sine-Gordon kink fluctuations, both phenomena subjected to Dirichlet boundary conditions. One or two delta wells will be also explored in order to describe absorbent plates, even though the wells lead to non unitary Quantum Field Theories.
\bigskip

{\bf Keywords:} One-dimensional scattering \and singular potentials \and Casimir effect \and sine-Gordon kink fluctuations.
\bigskip

{\bf PACS:} 03.65.Nk, 03.70.+k, 11.10.Lm
\end{abstract}

\section{Introduction}
\label{intro}
Our goal in this paper is to explore some one-dimensional scattering problems that are behind the scene in some
related (1+1)-dimensional scalar field theories. In particular, the scattering and bound state wave functions of the
quantum mechanical systems provide the one-particle states in the associated quantum field theories. In References \cite{Bordag:1998vs}, \cite{Bordag:2009zzd}, \cite{Milton:2004vy} and \cite{Milton:2004yy} it has been suggested that Dirichlet boundary conditions can be mimicked by two Dirac-$\delta$ barriers of infinite strength. In these References Green functions techniques has been used. We instead will apply one-dimensional scattering concepts, see  \cite{galpasqm}-\cite{Boyarev}, to address similar problems. We will also slightly depart from the scattering scenario to study how well defined unitary quantum field theories in a finite interval can be constructed when the $\delta$-walls are impenetrable. The framework developed in \cite{Symanzik:1981wd}, \cite{aim} and \cite{jmphd} to deal with quantum field theories defined on manifolds with boundary arise naturally from a much more physical point of view.

We will go further than this by applying a similar line of reasoning (technically more involved) to the analysis of sine-Gordon kink quantum fluctuations up to the one-loop level, see e.g. \cite{Alonso Izquierdo:2002rz}-\cite{CaveroPelaez:2009vi} to find different descriptions of this phenomenon but both akin to the framework of this paper. It is convenient to point out that in this later kind of problems (kink fluctuations) normal ordering is not enough to achieve ultraviolet renormalization but mass renormalization graphs must be taken into account. For the sake of brevity, we will not address this issue in this paper.

We have chosen this theme to celebrate M. Gadella 60th birthday because Manolo became a master in the study of this type of systems, see \cite{gadella}-\cite{gadella2}, in his long quest for understanding the very subtle nature of quantum resonances.
\section{Scalar field fluctuations.}

\subsection{Quantum scalar fluctuations and scattering problems in quantum mechanics}
The fluctuations of 1D free scalar fields on static classical backgrounds are governed by the following quadratic Lagrangian in the fields:
\[
{\cal L}=\frac{1}{2}\partial_\mu\Phi\partial^\mu\Phi-U(x)\Phi^2(x,t) \, \, \, , \quad \lim_{x\to\pm\infty} U(x)=0 \, \, , \, \, \int_{-\infty}^\infty \, dx \, U(x) < +\infty   \quad .
\]

We assume classical backgrounds as given by the function (or  generalized function) $U(x)$  that tend to zero at the two boundary points $x=\pm\infty$ of the spatial line enclosing a finite area {\footnote {$U(x)$ could also be interpreted as a variable square mass of the free quanta.}}. The Fourier components of the field $\Phi(t,x)=\int_{-\infty}^\infty \, \frac{dw}{2\pi}\, e^{i\omega t}\phi_\omega(x)$ satisfy the Sch$\ddot{\rm o}$dinger
equation:
\begin{equation}
 -\phi^{\prime\prime}_\omega(x)+U(x)\phi_\omega(x)=\omega^2 \phi_\omega(x) \qquad \quad \label{scfluc}
\end{equation}
such that the fluctuating normal modes are the solutions of the spectral problem (\ref{scfluc}). In general there
will be scattering solutions as well as bound states in this problem. Assuming that $\omega^2$ is definite positive
-$\omega^2=0$ modes do not contribute and $\omega^2<0$ modes are tachyonic- the uncertainty principle dictates that the contribution of these quantum fluctuations to the energy of the vacuum is of the form:
\begin{equation}
E_V=\sum \,  \omega - \sum \,  \omega_0= \sum_{j=1}^N \, \omega_j +\frac{1}{2}\int_{-\infty}^\infty \frac{dk}{2\pi} k\left[\frac{d\delta_+}{dk}+\frac{d\delta_-}{dk}\right] \qquad , \label{vacen}
\end{equation}
The sum of the bound state energies plus the integral due to the continuous part of the spectrum. We shall use throughout the paper the natural system of units where the Planck constant is equal to $2\pi$ and the speed of light is one: $\hbar=c=1$.

Even if the infrared divergences were tamed by choosing a very long normalization length $L$, ultraviolet divergences enter the game due to the infinite number of fluctuating modes. We subtract the contribution of the modes in absence of the background encoded in the spectral density of the problem (\ref{scfluc}) when the potential is zero, $\rho_{S_0}=\frac{L}{2\pi}$ , in order to cope with very energetic modes that do not feel the background. Thus, the spectral density measured with respect to the spectral density of the zero background case is given in terms of the derivative of the phase shifts with respect to the momentum:
\[
 \rho_S(k)-\rho_{S_0}=\frac{1}{4\pi}\left[\frac{d\delta_+}{dk}+\frac{d\delta_-}{dk}\right]
 \, \, .
\]
\subsection{Two Dirac $\delta$-potential and two-$\delta\oplus$P$\ddot{\rm o}$sch-Teller potential: the Casimir effect and kink fluctuations}
It has been suggested in the Literature \cite{Bordag:1998vs}-\cite{Milton:2004yy}-\cite{Milton:2004vy} that the Casimir effect between perfect conducting plates can be explained in this framework by considering a background given by two Dirac $\delta$ functions. In the interesting paper \cite{Bordag:1992cm} by Bordag and collaborators, for instance, the quantum vacuum interaction between two identical delta walls is studied for the cases of scalar and spinor fluctuations. The Lagrangian
\[
{\cal L}=\frac{1}{2}\partial_\mu\Phi\partial^\mu\Phi-\left(\alpha \delta(x+a)+ \beta \delta(x-a)\right)\Phi^2(x,t)
\]
mimics two homogeneous plates located at the points $x=\pm a$ with different penetrability related to the different wall strengths $\alpha$ and $\beta$. In order to use non-dimensional coordinates and coupling constants we perform the re-scalings:
\begin{equation}
  x+a\rightarrow \frac{1}{\Lambda}(x+a);\,t\rightarrow \frac{1}{\Lambda}t;\, \alpha\rightarrow\Lambda\alpha;\, \beta\rightarrow\Lambda\beta
\end{equation}
where $\Lambda$ is an small mass and $L=\frac{1}{\Lambda}>>a>0$ works as an infrared cutoff. Note that the Lagrangian density scales as ${\cal L}\rightarrow\Lambda^2
{\cal L}$.
\par
The spectral problem that comes out from the dimensionless Lagrangian density
\begin{equation}
\Delta \phi_\omega(x)=\left[-\frac{d^2}{dx^2}+\alpha \delta(x+a)+ \beta \delta(x-a)\right]\phi_\omega(x)=\omega^2\phi_\omega(x) \label{delta}
\end{equation}
and it is expressed in terms of non-dimensional fields, coordinates, coupling constants and $\delta$
functions will be solved in the next Section.

Another physical phenomenon related to 1D scattering processes is the surge of quantum fluctuations over classical kinks, see e.g. \cite{Alonso Izquierdo:2002rz}. We shall focus on the sine-Gordon model where the dynamics is determined by the Lagrangian density
\begin{equation*}
  {\cal L}=\frac{1}{2}\partial_\mu\Phi\partial^\mu\Phi+\frac{m^4}{\lambda} \left(\cos\left(\frac{\sqrt{\lambda}}{m}\Phi\right)-1\right)\quad .
\end{equation*}
Following \cite{rajaraman} we re-define fields, coordinates, and parameters to deal with non-dimensional quantities:
$\Phi\rightarrow\frac{m}{\sqrt{\lambda}}\Phi$; $x\rightarrow x/m$; $t\rightarrow t/m$.
The Lagrangian density scales as ${\cal L}\rightarrow \frac{m^4}{\lambda}\,{\cal L}$, where the non dimensional counterpart is
\[
{\cal L}=\frac{1}{2}\partial_\mu\Phi\partial^\mu\Phi+\cos\Phi-1
\]
and the static configuration $\phi_K(x)=4\arctan e^x$
is the famous stable kink solution of the sine-Gordon equation. The shift from the kink field $\Phi(t,x)= \phi_K(x)+H(t,x)$ shows the Lagrangian for the \lq\lq Higgs" boson field $H(t,x)$ in the solitonic background:
\begin{eqnarray*}
{\cal L}&=&\left\{\frac{1}{2}\left(\frac{\partial H}{\partial t}
\frac{\partial H}{\partial t}-\frac{\partial H}{\partial x}
\frac{\partial H}{\partial x}\right)+\left(1-\frac{2}{\cosh^2x}\right)\sum_{n=1}^\infty (-1)^n\frac{H^{2n}(t,x)}{(2n)!}\right.\\
&+&\left.2\frac{\tanh x}{\cosh x}\sum_{n=1}^\infty (-1)^n\frac{H^{2n+1}(t,x)}{(2n+1)!}\right\}
\end{eqnarray*}
The small kink fluctuations, however, are ruled by a Lagrangian quadratic in the field, i.e., by keeping only the $n=1$ term in the expansions above:
\[
{\cal L}^{(2)}= \frac{1}{2} \left\{\frac{\partial H}{\partial t}
\frac{\partial H}{\partial t}-\frac{\partial H}{\partial x}
\frac{\partial H}{\partial x}-(1-\frac{2}{\cosh^2x})H^2(t,x)\right\} .
\]
The Fourier modes satisfy now the Schr$\ddot{\rm o}$dinger equation
\begin{equation}
\Delta \phi_\omega(x)=\left[-\frac{d^2}{dx^2}+1-\frac{2}{\cosh^2x}\right]\phi_\omega(x)=\omega^2\phi_\omega(x)
\label{SPT}
\end{equation}
although the energy of the kink fluctuations is given by one formula like the formula of the vacuum energy only at the one-loop level. Moreover, because there are many vertices in the Lagrangian it will be necessary normal ordering the Hamiltonian to control all the ultraviolet divergences. We shall also add two $\delta$-walls to take into account the effect of both one external -the two $\delta's$- and one solitonic background -the P$\ddot{\rm o}$sch-Teller
potential- mixed together. Therefore, a new, interesting, scattering problem will be studied.

We finish this Section by showing the meson Green function in the kink background calculated according to standard techniques in QFT \cite{CaveroPelaez:2009vi}:
\begin{eqnarray*}
 && \langle 0;K|T\left(\hat{H}(x^\mu)\hat{H}(y^\mu)\right)|K; 0 \rangle=\\\
 &&=\int\frac{d\omega}{2\pi}\int\frac{dk}{2\pi}\cdot \frac{e^{-i\omega(x^0-y^0)+ik(x^1-y^1)}}{(1+k^2)(\omega^2-k^2+1+i\varepsilon)}P_1(x^1,k)P_1(y^1,-k)
\end{eqnarray*}
where $P_1(z,k)=\tanh z+i k$ is the first-order Jacobi polynomial in $\tanh z$.

\section{The two-$\delta$ potential}

 We start this Section by stating the following CAVEAT:
 Even though the one-particle scattering  over attractive $\delta$'s is a well defined quantum mechanical problem, the associated $1+1$-dimensional Quantum Field Theory suffer from pathologies. The QFT model is non unitary when the quantum fluctuation operator is non positive, i. e., when bound states of negative energy appear in the spectrum of the one-particle Hamiltonian operator. Nevertheless, absorbent plates and/or creation/anihilation of particles can be modeled by $\delta$ wells and, thus, we shall not completely rule out attractive $\delta$ potentials.

\subsection{Scattering by the two-$\delta$ potential}

We choose the potential in the form: $U(x)=\alpha \delta(x+a)+ \beta \delta(x-a)$. To build scattering wave functions we divide the real line in three zones:

\[
 1) \, \, {\rm Zone} \, \text{II} : x < - a \, . \quad , \quad 2) \, \, {\rm Zone} \, \text{I}: -a < x < a \, . \quad , \quad
 3) \, \, {\rm Zone}\, \text{III}: x > a \, .
\]

The ``diestro'' (righthanded) - particles incoming from the left- scattering wave functions are of the form:
\[
\psi_r(x)=
\begin{cases}
 e^{-i k x} \rho _r+e^{i k x} & \, , \, \, \, \, x\in \text{II} \\
 A_r e^{i k x}+B_r e^{-i k x} & \, , \, \, \, \, x\in \text{I} \\
 e^{i k x} \sigma _r & \, , \, \, \, \, x\in \text{III}
\end{cases}
\]
Clearly, the functions above are eigen-functions of the Schr$\ddot{\rm o}$dinger operator in each zone. In order
to become solutions in the whole real line, they must be sewed by demanding continuity of the wave function $\psi$ and setting the finite step discontinuity of $\psi^\prime$ at the points $x=\pm a$:
\begin{eqnarray*}
&& \psi(\pm a_<)=\psi(\pm a_>) \, \, \, , \quad \psi^\prime(\pm a_<)-\psi^\prime(\pm a_>)=\lim_{\delta\to 0}
\int_{\pm a-\delta}^{\pm a+\delta}\, dx \, U(x)\psi(x) \, \, \Rightarrow \\ && \, \, \Rightarrow \psi^\prime(-a_<)-\psi^\prime(-a_>)=\alpha \psi(-a)
\, \, \, , \quad \psi^\prime(a_<)-\psi^\prime(a_>)=\beta \psi(a)
\end{eqnarray*}
These four equations are sufficient to fix the transmission and reflection amplitudes (``diestras''),
$\sigma_r$, $\rho_r$, as well as the amplitudes of the wave functions $A_r$, $B_r$ between the walls/wells:
\begin{eqnarray*}
   \rho_r &=& -\frac{i e^{-2 i a k} \left(\beta  e^{4 i a k} (2 k-i \alpha )+\alpha(2k+i \beta )\right)}
   {\bigtriangleup(k)}\\
   A_r &=& \frac{2 k (2 k+i \beta )}{\bigtriangleup(k)}\, \, \, , \, \, \,
   B_r = -\frac{2 i k \beta  e^{2 i a k}}{\bigtriangleup(k))}\\ \sigma_r &=& \frac{4 k^2}{\bigtriangleup(k)} \, \, \, , \, \, \, \bigtriangleup(k)=\alpha\beta\left(-1+e^{4 i a k}\right)+4 k^2+2 i k (\alpha +\beta)
\end{eqnarray*}

For particles coming from the right the \lq\lq zurdo" (lefthanded) scattering wave functions have the form:
\[
\psi_l(x)=\begin{cases}
 e^{-i k x} \sigma _l & \, , \, \, \, \, x\in \text{II} \\
 A_l e^{i k x}+B_l e^{-i k x} & \, , \, \, \, \, x\in \text{I} \\
 e^{i k x} \rho _l+e^{-i k x} & \, , \, \, \, \, x\in \text{III}
\end{cases}
\]
Arguing exactly in the same way as in the \lq\lq diestro scattering we find the transmission and reflection amplitudes (\lq \lq zurdas'') as well as the intertwining amplitudes:
\begin{eqnarray*}
   \sigma_l &=& \frac{4 k^2}{\bigtriangleup(k)}\\
   A_l &=& -\frac{2 i k \alpha  e^{2 i a k}}{\bigtriangleup(k)}\, \, \, , \, \, \,
   B_l = \frac{2 k (2 k+i \alpha )}{\bigtriangleup(k)}\\
   \rho_l &=& -\frac{i e^{-2 i a k}\left(\alpha  e^{4 i a k} (2 k-i \beta )+\beta
   (2 k+i \alpha )\right)}{\bigtriangleup(k)}
\end{eqnarray*}

From these data we obtain the 2 $\delta$ scattering matrix: ${\bf S}[k,\alpha,\beta, a]=\left(
\begin{array}{cc}
 \sigma _r & \rho _l \\
 \rho _r & \sigma _l
\end{array}
\right)$ .

The phase shifts, the eigenvalues of the scattering matrix, and the spectral density are consequently defined in the form:
\[
e^{2 i \delta_\pm}= \sigma \pm \sqrt{\rho_l\rho_r} \qquad , \qquad \rho_S(k)=\frac{1}{2\pi}
\frac{d (\delta_++\delta_-)}{dk}+\rho_{S_0}
\]
Note that, due to the invariance of the Schr$\ddot{o}$dinger equation under time inversion, $\sigma_r=\sigma_l$,
whereas $\rho_r=\rho_l$ only for $x\to -x$ invariant (even) potentials.

\subsection{Impenetrable barriers: Dirichlet boundary conditions}
 We now consider the double $\alpha\to +\infty$ and $\beta\to +\infty$ limit. For arbitrary values of the momentum only the reflection amplitudes $\rho_r$ and $\rho_l$ are non zero because the $\delta$-barriers become completely opaque. There are, however, values of $k$ such that $\rho_r=\rho_l$ is singular and $A_r=B_l$, $B_r=A_l$ are non null but finite. These momenta arise in a situation where the zones II and III are completely disconnected between them and from the zone I. This is no one scattering scenario, rather, the waves remain confined in the finite interval $[-a,a]$ subjected to some boundary conditions to be specified.

When $k$ takes the special values mentioned above, the quantum mechanical system that arises in zone I is a consistent system in the sense of \cite{jmphd,aim}. In this case the Laplace-Beltrami operator restricted to zone I is self-adjoint, and the Hamiltonian unitary. Because in this limit the Hamiltonian is definite positive, a unitary Quantum Field Theory can be constructed in the zone I bounded region \cite{jmphd}. Necessary conditions for having consistent quantum mechanical and  quantum field theoretical systems in zone I are encoded in the boundary behaviour of the wave functions.

 In order to do that, we denote
 \[
 \Delta(k)=\alpha  \beta  \Delta_2(k)+4 k^2+2 i k (\alpha+\beta) \, \, \, , \quad \Delta_2(k)=\left(-1+e^{4 i a k}\right) \, \, \, .
 \]
\[
\Delta_2(k)=e^{2iak}2i\sin (2ak)=\frac{1}{2}e^{2iak}h_D^{(2a)}(k)
\]
is, up to a phase and a factor of 2, the spectral function for Dirichlet boundary conditions derived in
\cite{jmphd} within the framework of the \cite{aim} formalism to construct unitary quantum field theories
in manifolds with boundary (see also the references \cite{asmajo,Asorey:2008xt,Asorey:2007rt} where this point of view is further developed).

In fact, the zeroes of $h_D^{(2a)}(k)$
\[
h_D^{(2a)}(k_n)=0 \qquad , \qquad k_n=\frac{\pi}{2a} n \quad  , \quad n\in{\mathbb Z}
\]
are the only values of the momentum for which the $A_r$, $B_r$, $A_l$, $B_l$ amplitudes
do not go to zero in the $\alpha , \beta \to +\infty$ limit but they become:
\[
  A_r(k_n)=1/2=B_l(k_n) \quad , \quad B_r(k_n)=-\frac{e^{2 i a k_n}}{2}=A_l(k_n)
\]
For these allowed momentum values the zone I wave functions take the form
\[
\psi(x,k_n)=\psi_r(x,k_n)=-e^{2iak_n}\psi_l(x,k_n)=\frac{1}{2}\left(e^{ik_nx}-e^{2iak_n}e^{-ik_nx}\right)
\]
in the ultra-strong limit, collapsing the \lq\lq diestro" and \lq\lq zurdo" solutions in a single
solution that we denote as $\psi(x,k_n)$. There are two possibilities:
\begin{itemize}
  \item If n is even, the wave function is:
  \[
    \psi(x,k_n)=\sin \frac{\pi}{2a}n x
  \]
  Because $\sin(- \frac{\pi}{2a}n x)=-\sin \frac{\pi}{2a}n x$ only the even positive integers,
  $n=2,4,6,8,\cdots$, provide linearly independent wave functions. Moreover, for n even Dirichlet boundary
  conditions on the $[-a,a]$ interval are satisfied: $\psi(-a,\frac{\pi}{2a}n)=\psi(a,\frac{\pi}{2a}n)=0$. The
  spectral condition that characterizes the odd states is
  \begin{equation*}e^{2 i a k_n}-1=0\end{equation*}
  \item If n is odd, the wave function is:
  \[
    \psi(x,k_n)=\cos \frac{\pi}{2a}n x
  \]
  Because $\cos(- \frac{\pi}{2a}n x)=\cos \frac{\pi}{2a}n x$ only the odd positive integers, $n=1,3,5,7,\cdots $,
  provide linearly independent wave functions. Moreover, for n odd Dirichlet boundary conditions on the $[-a,a]$
  interval are satisfied: $\psi(-a,\frac{\pi}{2a}n)=\psi(a,\frac{\pi}{2a}n)=0$. The even states, are
  characterized by the spectral condition
  \begin{equation*}e^{2 i a k_n}+1=0\end{equation*}
\end{itemize}

Once we know the spectrum and its corresponding states, it is easy to compute the dimensionless Casimir energy between two perfect conducting plates using zeta function techniques. The non regularized expression of the vacuum energy due to fluctuations of a real scalar field is the divergent series:
\begin{equation}
  E_d={1\over 2}\sum_{n=1}^\infty\left(n^2\pi^2/(2a)^2\right)^{1/2}\quad .
\end{equation}
We regularize the series by means of the zeta regularization prescription:
\begin{equation}
  E_d(s)={1\over 2}\sum_{n=1}^\infty\left(n^2\pi^2/(2a)^2\right)^{-s}={1\over 2}\left({\pi\over 2a}\right)^{-2s}\zeta(2s)  \, , \quad  s\in\mathbb{C}. \label{riemmz}
\end{equation}
The convergence domain for $\zeta(2s)$ understood as the series in (\ref{riemmz}) is ${\rm Re}(2s)\geq 1$. Standard analytical continuation of this function $\zeta(2s)$ to all the complex $s$-plane provides the Riemann zeta function, a meromorphic function with a single pole at $2s=1$. The physical limit $s\rightarrow-1/2$ is a regular point and $\zeta(-1)$ gives the Casimir energy for the Dirichlet boundary conditions:
\begin{equation}
  E_d=\frac{\pi}{4}\zeta(-1)=-\frac{\pi}{48 a}\;\label{zetadir}.
\end{equation}

\section{The two-$\delta$ $\oplus$ P$\ddot{\rm o}$sch-Teller potential}

\subsection{Scattering by the two-$\delta$ $\oplus$ P$\ddot{\rm o}$sch-Teller potential}

The potential describing the propagation of mesons moving in a sine-Gordon kink background plus two-$\delta$ potentials is:
\[
U(x)=\alpha \delta(x+a)+ \beta \delta(x-a)+1-(\theta(a-x)\cdot\theta(a+x))\frac{2}{{\rm cosh}^2 x}
\]
Under the standard re-scaling used for the sine-Gordon model, the delta strengths are re-scaled as $\alpha\rightarrow m\alpha;\,\beta\rightarrow m\beta$.
The scattering solutions are build in the same form as in the previous Section but one must replace plane waves
by plane waves times first-order Jacobi polynomials (solutions of the P$\ddot{\rm o}$sch-Teller spectral problem) in the
zone I:
\[
\psi_r=\begin{cases}
 e^{-i k x} \rho _r+e^{i k x} & \, , \ \, \, \, x\in \text{II} \\
 A_r f_k(x)+B_r f_{-k}(x) & \, , \, \, \, \, x\in \text{I} \\
 e^{i k x} \sigma _r & \, , \, \, \, \, x\in \text{III}
\end{cases} \, \, \, , \qquad f_k(x)=e^{i k x}\left({\rm tanh} x-i k\right)
\]
Identical continuity/discontinuity conditions at $x=\pm a$ as before provide
the (\lq\lq diestras'') transmission and reflection amplitudes as well as the amplitudes of the wave functions $A_r$, $B_r$ in zone I: \, \,
\begin{eqnarray*}
   \rho_r &=& \frac{1}{\Delta^{(K)}(k)}\left(e^{2 i a k} \left(s^2+\alpha  (t-i k)\right) \left(s^2+(k+i t) (2 k-i
   \beta )\right)\right.\\
   &-&\left.e^{-2 i a k}
   \left(s^2+(k-i t) (2k+i \alpha )\right) \left(s^2+\beta  (t+i k)\right)\right)\\
   A_r &=& \frac{2 k e^{2 i a k} \left(\alpha  (k+i t)+i s^2\right)}{\Delta^{(K)}(k)}\, \, \, , \, \, \,
   B_r = -\frac{2 i k \left(s^2+(k-i t) (2 k+i \alpha )\right)}{\Delta^{(K)}(k)}\\
   \sigma_r &=& \frac{4 \left(k^4+k^2\right)}{\Delta^{(K)}(k)}
\end{eqnarray*}
where
\begin{eqnarray*}
  \Delta^{(K)}(k)&=&\left(s^2+(k-i t) (2 k+i \alpha )\right) \left(s^2+(k-i t) (2 k+i \beta )\right)\\
  &-&e^{4 i a k} \left(s^2+\alpha  (t-i k)\right)
   \left(s^2+\beta  (t-i k)\right)
\end{eqnarray*}
and we have used the abbreviations: $s\equiv \text{sech(a)}$, \, \,  $t\equiv\text{tanh(a)}$.

Simili modo, the \lq\lq zurdo'' scattering solutions are:
\[
\psi_l(x)=\begin{cases}
 e^{-i k x} \sigma _l & \, , \, \, \, \, x\in \text{II} \\
 A_l f_k(x)+ B_l f_{-k}(x) & \, , \, \, \, \, x\in \text{I} \\
 e^{i k x} \rho _l+e^{-i k x} & \, , \, \, \, \, x\in \text{III}
\end{cases}
\]

Properly glued we find the transmission and reflection amplitudes (\lq\lq  zurdas \rq\rq) as well as
the $A_l$, $B_l$ amplitudes in the zone I:
\begin{eqnarray*}
   \rho_l &=& \frac{1}{\Delta^{(K)}(k)}\left(e^{2 i a k} \left(s^2+\beta  (t-i k)\right) \left(s^2+(k+i t) (2 k-i
   \alpha )\right)\right.\\
   &-&\left.e^{-2 i a k}
   \left(s^2+(k-i t) (2k+i \beta )\right) \left(s^2+\alpha  (t+i k)\right)\right)\\
   A_l &=& \frac{2 k e^{2 i a k} \left(\beta  (k+i t)+i s^2\right)}{\Delta^{(K)}(k)}\, \, \, , \, \, \, \,
   B_l = -\frac{2 i k \left(s^2+(k-i t) (2 k+i \beta )\right)}{\Delta^{(K)}(k)}\\
   \sigma_r &=& \frac{4 \left(k^4+k^2\right)}{\Delta^{(K)}(k)}
\end{eqnarray*}
From all this information we obtain the scattering matrix, the phase shifts and the spectral density:
\[
S=\left(
\begin{array}{cc} \sigma_r & \rho_l
\\ \rho_r & \sigma_l
\end{array}\right)
\]
\[
e^{2 i \delta_\pm}= \sigma \pm \sqrt{\rho_l\rho_r} \qquad , \qquad \rho_S(k)=\frac{1}{2\pi}
\frac{d (\delta_++\delta_-)}{dk}+\rho_{S_0}
\]

\subsection{Impenetrable barriers: Dirichlet boundary conditions}
Like in the two-$\delta$ problem we search now for the special values of the momentum for which non-zero
wave functions survive between the plates when the walls become impenetrable: $\alpha,\beta\to +\infty$, .
Id est, we address a problem where Dirichlet boundary conditions are imposed on a P$\ddot{\rm o}$sch-Teller potential. In this limit the potential is even and the right and left solutions coincide, so it is only necessary to study the right-handed case\footnote{We will only study the regime of long separation between the two delta walls. There is a critical value $a_c$ for the separation between deltas that distinguishes between the short and long distance regimes where the spectra are qualitatively different (see \cite{jmgjmmc}, in preparation).}. We suppress accordingly the right and left sub-indices along this subsection

We write the right-handed amplitudes in zone I in the form
\[
  A(k)= \frac{2 i k \left(\text{s}^2+(2
   k+i \alpha ) (k-i \text{t} )\right)}{\Delta^{^{(K)}}(k)} \, \, \, , \, \, \,
   B(k)= -\frac{2 k e^{2 i a k} \left(\alpha  (k+i \text{t})+i
   \text{s}^2\right)}{\Delta^{^{(K)}}(k)}.
\]
The amplitudes $A$ and $B$ are non null and finite in the ultra-strong limit if and only if the term proportional to $\alpha\beta$ in the denominator ($\Delta^{^{(K)}}(k)$)
\begin{equation}
   \Delta^{^{(K)}}_2(k)=\left(e^{2 i a k} (k+i t)- (k-i t)\right) \left(e^{2 i a k} (k+i t)+ (k-i t)\right)
   \label{facd}
\end{equation}
is zero. This happens for those momenta that annihilate this term: $ \Delta^{^{(K)}}_2(k_n)=0$. For these values of the momentum the amplitudes become
\[
  \lim_{\alpha,\beta\to\infty}A(k_n)=\frac{k_n}{\Delta^{^{(K)}}_1(k_n)}(k_n-i t)\quad , \quad
  \lim_{\alpha,\beta\to\infty}B(k_n)=\frac{k_n}{\Delta^{^{(K)}}_1(k_n)}e^{2 i a k_n}(k_n+it) \qquad ,
\]
when the walls are impenetrable (i. e. in the limit $\alpha,\,\beta\to\infty$).
Here $2\Delta^{^{(K)}}_1(k)$ is the first-order term in $\alpha$ of $\left.\Delta^{^{(K)}}(k)\right|_{\beta=\alpha}$:
\begin{equation*}
  \Delta^{^{(K)}}_1(k)=s^2 \left(t \left(-1+e^{4 i a k}\right)-i k \left(3+e^{4 i a k}\right)\right)-
  2 i k \left(k^2-2 i k t-1\right)
\end{equation*}

\begin{figure}[h]
  \centerline{\includegraphics[height=3.5cm]{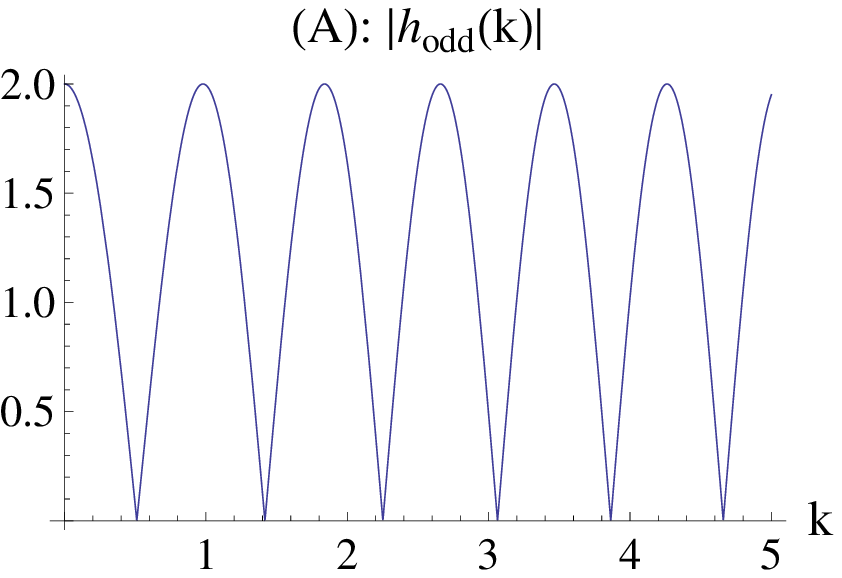}\quad\includegraphics[height=3.5cm]{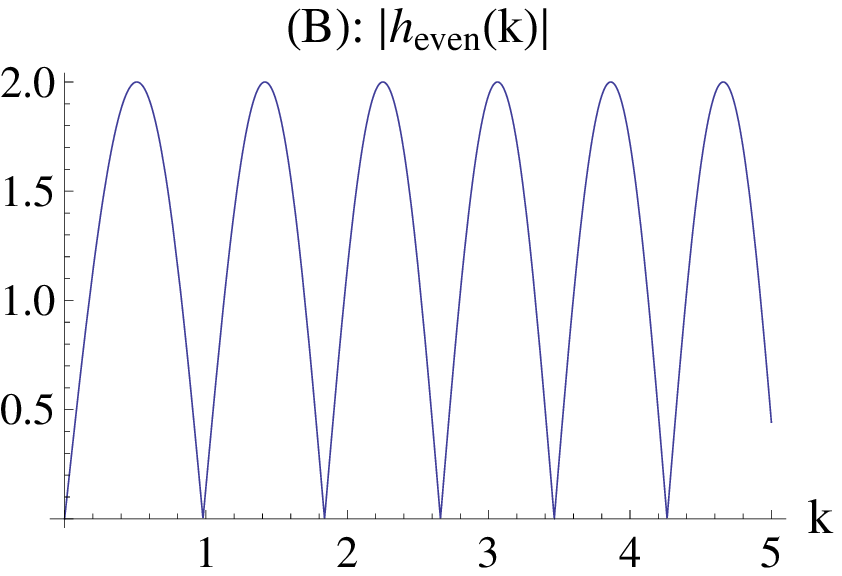}}
\caption{Representation of $|h_{odd}(k)|$ (A), and $|h_{even}(k)|$ (B) as continuous functions of the momentum $k$, for $a=4$. The zeroes of $|h_{odd}(k)|$ (A), and $|h_{even}(k)|$ (B), define the spectrum of odd and even states respectively. The lowest energy level corresponds to the first non null zero of $|h_{odd}(k)|$, as it can be seen immediately comparing graphics (A) and (B). This does not mean that the ground state is odd. In fact,
it is even but corresponds to a purely imaginary root of $|h_{even}(k)|$, see later. }
\label{spectra}
\end{figure}

\begin{figure}[h]
  \centerline{\includegraphics[height=3.8cm]{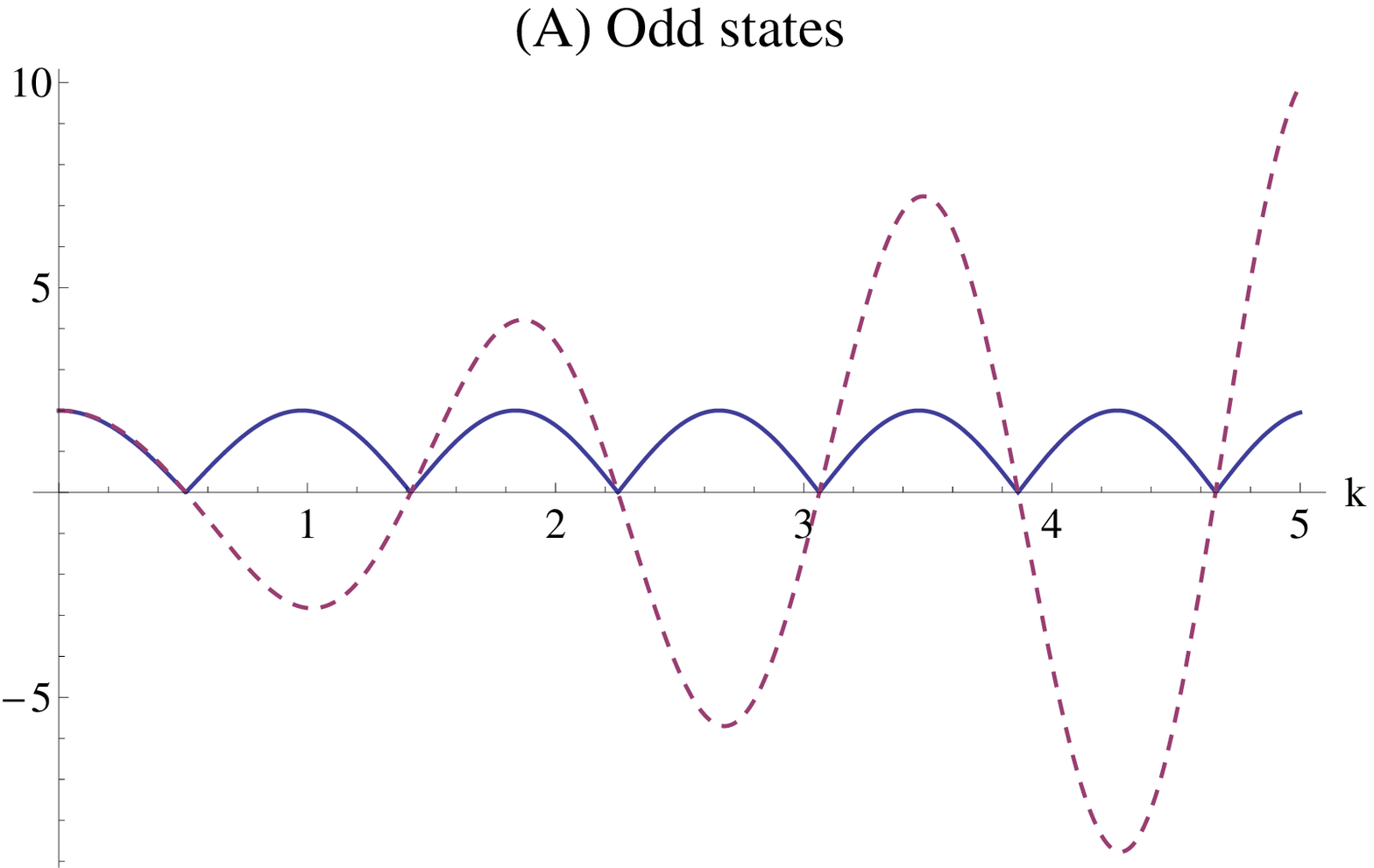}\quad\includegraphics[height=3.8cm]{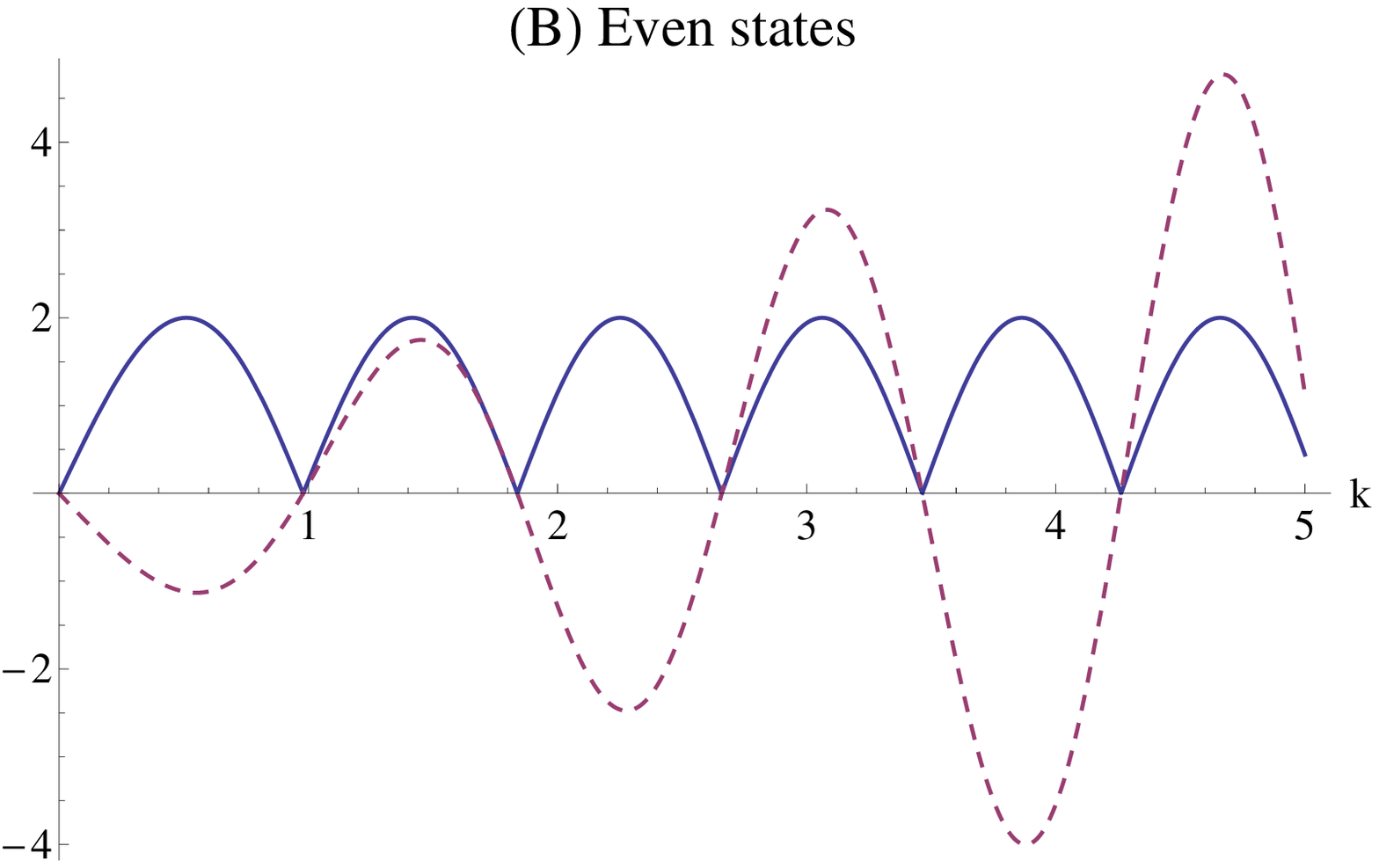}}
\caption{(A): Representation of  $|h_{odd}(k)|$ (continuous line) and $\psi^{(-)}_k(a)$ (dashed line), both as functions of k. (B): Representation of  $|h_{even}(k)|$ (continuous line) and $\psi^{(+)}_k(a)$ (dashed line), both as functions of k. In both cases, it has been taken $a=4$}
\label{statesbv}
\end{figure}

Because of the factorization of $\Delta^{^{(K)}}_2(k)$ given in (\ref{facd}) there are two possibilities that annihilate $\Delta^{^{(K)}}_2$. Each possibility gives rise to one-half of the spectrum:
\begin{itemize}
  \item The first factor is zero if the transcendent equation
  \begin{equation}
     h_{odd}(k_n)=e^{2 i a k_n} (k_n+i t)- (k_n-i t)=0 \quad , \label{osmom}
  \end{equation}
  is satisfied.The equation (\ref{osmom}) has infinite solutions, see figure \ref{spectra}
  (A). The sub-index $n$ labels the $n$th solution of (\ref{osmom}). If $k_n$ solves
  (\ref{osmom}) $A(k_n)=B(k_n)$ in the ultra-strong limit  and the associated eigenfunctions are
  odd in $x\to -x$:
  \begin{equation}
    \psi^{(-)}_{k_n}(x)=\frac{2 k_n}{\Delta^{^{(K)}}(k_n)}(k_n-it)\left(k_n \sin (k_n x)+\tanh (x) \cos (k_n
    x)\right)\label{oswfun}
  \end{equation}
   It is also easy to check from figure \ref{statesbv} (A) that for the momenta $k_n$ solving (\ref{esmom}) these
   eigen-functions are null at $x=\pm a$ $\psi^{(-)}_{k_n}(-a)=\psi^{(-)}_{k_n}(a)=0$, complying with Dirichlet
   boundary conditions.

  \item If the momenta satisfy the transcendent equation
  \begin{equation}
   h_{even}(k_n)=e^{2 i a k_n} (k_n+i t)+ (k_n-i t)=0 \label{esmom}
  \end{equation}
  they also annihilate the quadratic coefficient and provide non-null but finite values for the amplitudes in the
  ultra-strong limit. In this case, however, we find that $A(k_n)=-B(k_n)$ when $k_n$ solves (\ref{esmom}) (see
  figure \ref{spectra} (B)). Therefore, the corresponding eigen-functions are even in $x\to -x$:
  \begin{equation}
    \psi^{(+)}_{k_n}(x)=\frac{2 i k_n}{\Delta^{^{(K)}}(k_n)}(k_n-i t)\left(k_n \cos (k_n x)-\tanh (x) \sin (k_n
    x)\right)\label{eswfun}
  \end{equation}
  Again, Dirichlet boundary conditions $\psi^{(+)}_{k_n}(-a)=\psi^{(+)}_{k_n}(a)=0$ are satisfied if $k_n$ is a solution of (\ref{esmom}), as it can
  be seen in figure \ref{statesbv} (B).
\end{itemize}

In both cases we need only to take into account the positive solutions $k_n>0$ because the exchange $k_n\to -k_n$
only gives the same function multiplied by a complex number.

 It remains to explore the pure imaginary roots of the $h_{odd}$ and $h_{even}$ spectral functions \footnote{We stress again that both the real and the imaginary roots that we are discussing arise in the long separation regime. We shall analyze the short separation spectrum in a forthcoming publication \cite{jmgjmmc}.}

\begin{figure}[h]
  \centerline{\includegraphics[height=3.8cm]{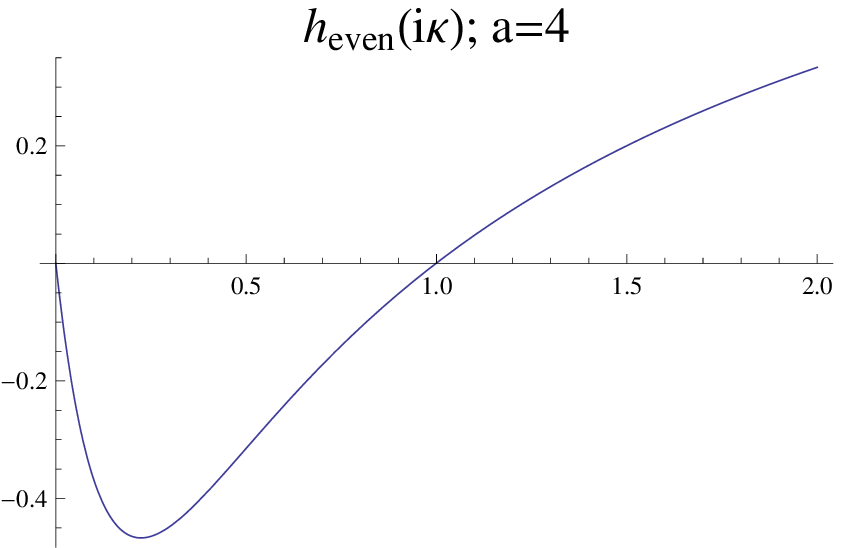}\quad\includegraphics[height=4cm]{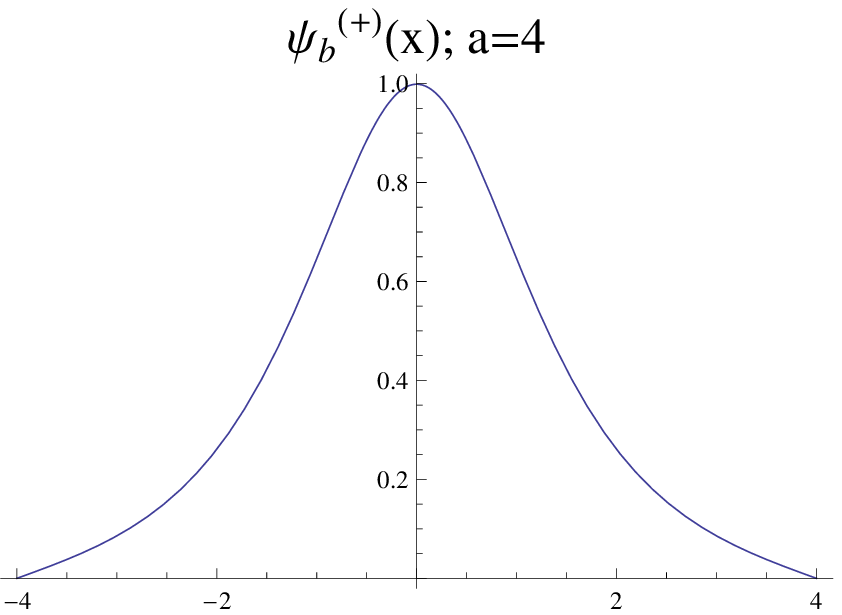}}
\caption{Left: there is only one solution for $h_{even}(i\kappa)=0$ with $a=4$ (long distance regime). Right: the wave function of the corresponding state is represented between $x=-a$ and $x=a$, for $a=4$.}
\label{bstate}
\end{figure}

\begin{itemize}
  \item The odd spectral function evaluated at $k=i \kappa$
  \begin{equation*}
    h_{odd}(i\kappa_n)=i\left(e^{-2 a \kappa_n}(\kappa_n+t)-(\kappa_n-t)\right),
  \end{equation*}
  has only one root for any value of the distance $a$: $\kappa=1$. Replacing $k=i$, however, in the formula \ref{oswfun} we obtain $\psi^{(-)}_i=0$ which is not a physical state.
  \item The even spectral function evaluated at $k=i\kappa$
  \begin{equation*}
    h_{even}(i\kappa_n)=i\left(e^{-2 a \kappa_n}(\kappa_n+t)+(\kappa_n-t)\right),
  \end{equation*}
  has one root in the long distance regime, see Figure \ref{bstate}. The value of this root can be computed numerically and it is given for, e.g., $a=4$ by $\kappa_b=0.9986$. Plugging $k=i \kappa_b$ in the expression \ref{eswfun}, we obtain the even ground state wave function plotted in Figure \ref{bstate} where one can see that it satisfies Dirichlet boundary conditions. In the $a\to\infty$ limit this  ground state becomes the well known bound state of the transparent P$\ddot{\rm o}$sch-Teller potential, i.e., $\lim_{a\to\infty}\kappa_b=1$, whereas the other states go to scattering states.
\end{itemize}

The short distance regime will be studied carefully in \cite{jmgjmmc}. The numerical value that separates both regimes is given by $a_c=1.1996786$. When $a<a_c$, the imaginary  root state disappears but the first real root state, the ground state, is even.

\section{Bound states, virtual states, and resonances}

 We devote this last Section to the comparison of the two different scattering problems. In particular, we are interested in knowing how bound states, virtual (anti bound) states, and resonances arise in 2 $\delta$ wells/walls as compared with 2 $\delta$ wells/walls plus PT wells. According to very well established theorems (see  e.g. \cite{galpasqm}) bound states are poles of the transition amplitude of the form $k=i\kappa$, virtual states, poles of the form $\kappa\in{\mathbb R}^+$, $k=-i\kappa$, $\kappa<<1$, and, resonances poles of the form  $k=\pm\gamma-i\kappa$, $\gamma>0$, $\kappa<<1$, in the $k$-complex plane.

We will focus only in the case of walls of the same strength $\alpha=\beta$ (wells, if $\alpha$ is negative).
The poles of $\sigma(k)$ are the zeros in the $k$ complex plane of the denominator $\Delta (k)$ that we write in terms of the Jost functions:
\begin{equation}
\Delta(k)=4J_0(k)J_1(k) , \quad J_0(k)=k+i\alpha e^{ika}\cos ka, \quad J_1(k)=k+\alpha e^{ika}\sin ka. \label{Jost}
\end{equation}
In the kink (P$\ddot{\rm o}$sch-Teller) case matters are identical conceptually, and in the search for bound/antibound states and resonances we look for the zeroes of the denominator of the transmission amplitude in the $k$-complex plane
of the same form as above:
\begin{eqnarray}
\Delta^{^{(K)}}(k)&=&4J_0^K(k)J_1^K(k)\Rightarrow\\ J_0^K(k)&=&\left(s^2+(2k+i\alpha)(k-it)\right)+e^{2aik}\left(s^2+\alpha(t-ik)\right)\nonumber \\
J_1^K(k)&=&\left(s^2+(2k+i\alpha)(k-it)\right)-e^{2aik}\left(s^2+\alpha(t-ik)\right)\label{Jost1} \qquad .
\end{eqnarray}
Therefore, the analysis of the intersection of the curves ${\rm Re}\Delta(k)=0$ with ${\rm Im}\Delta(k)=0$ and ${\rm Re}\Delta^{^{(K)}}(k)=0$ with ${\rm Im}\Delta^{^{(K)}}(k)=0$ will inform us about
the bound/antibound states and resonances of our scattering problems. {\footnote{Only the bound states count in the Casimir energies.}}

\begin{figure}[h]
  \centerline{\includegraphics[height=6cm]{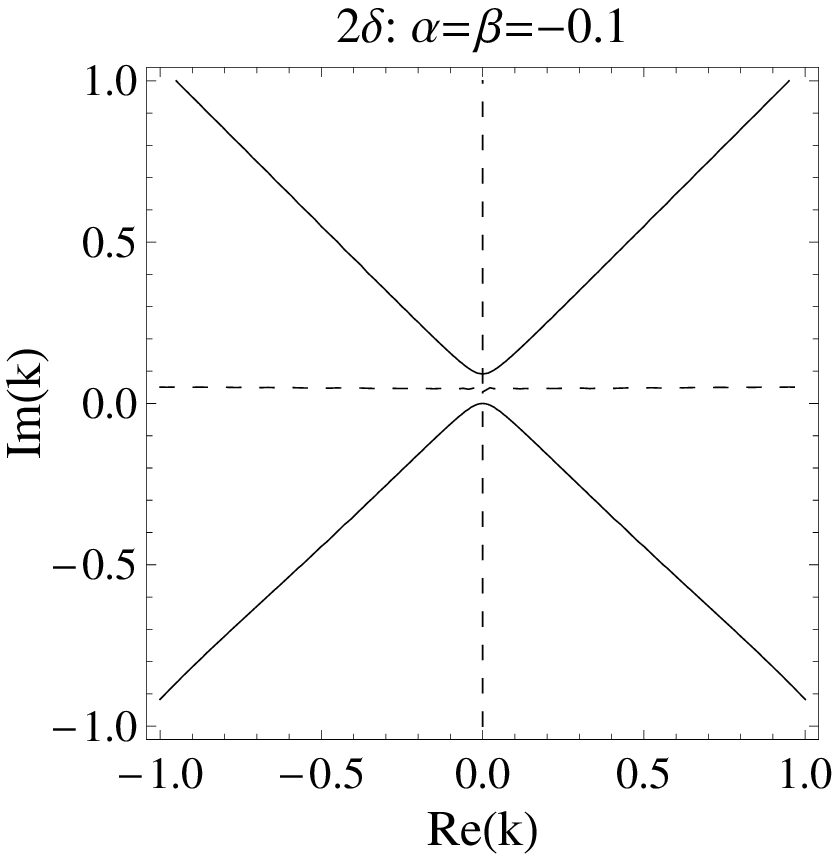}\quad\includegraphics[height=6cm]{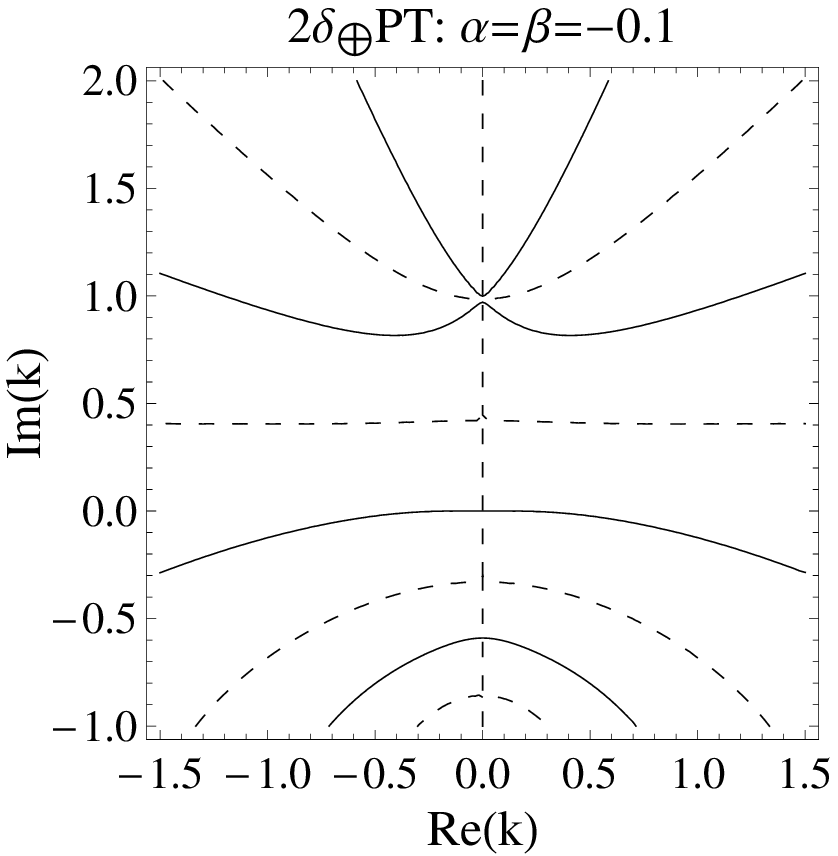}}
\caption{Plot of the curves: ${\rm Re}\Delta(k)=0$ (continuous lines) and
${\rm Im}\Delta(k)=0$ (dashed lines) (left). Plot of the curves: ${\rm Re}\Delta^{^{(K)}}(k)=0$ (continuous lines) and ${\rm Im}\Delta^{^{(K)}}(k)=0$ (dashed lines) (right). Weakly attractive case.}
\label{aa01}
\end{figure}

In Figure \ref{aa01} we see that, for a weakly attractive well $\alpha=\beta=-0,1$, these curves (continuous and dashed, respectively the zero locus of ${\rm Im}\Delta$ and ${\rm Re}\Delta$) intersect at one positive imaginary point in the two-$\delta$ potential and at two imaginary points in the two-$\delta$ plus P$\ddot{\rm o}$sch-Teller system. This means that there is a bound state in the first case and two bound states in the kink case. It seems that the very well known bound state $k=i$ that traps one meson to the kink pushes up the modulus of the momentum of the single bound state of the two-$\delta$ potential.

If the wells are more attractive, setting e.g. $\alpha=\beta=-2$, the number of bound states increases. In the
next Figure \ref{aa2} it can be seen that continuous and dashed lines intersect at the positive imaginary axis at two points for the two-$\delta$ well and at three points for the two-$\delta$ $\oplus$ PT well. Again, the PT potential add a bound state to those existing in the two-$\delta$ potential pulling upwards the lower momentum bound state.
\begin{figure}[h]
  \centerline{\includegraphics[height=6cm]{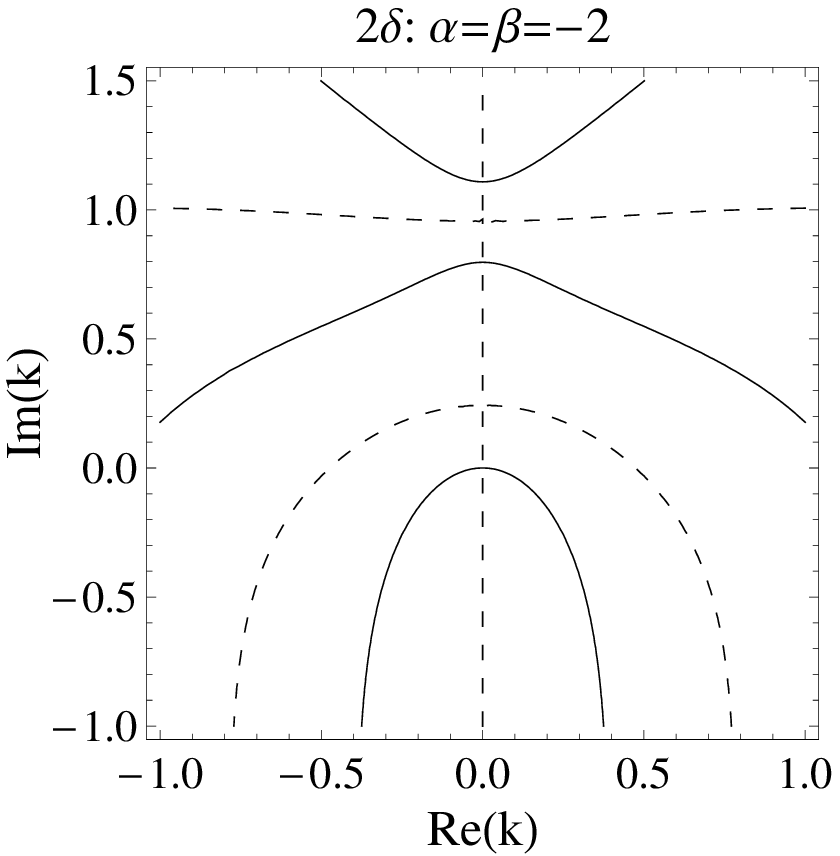}\quad\includegraphics[height=6cm]{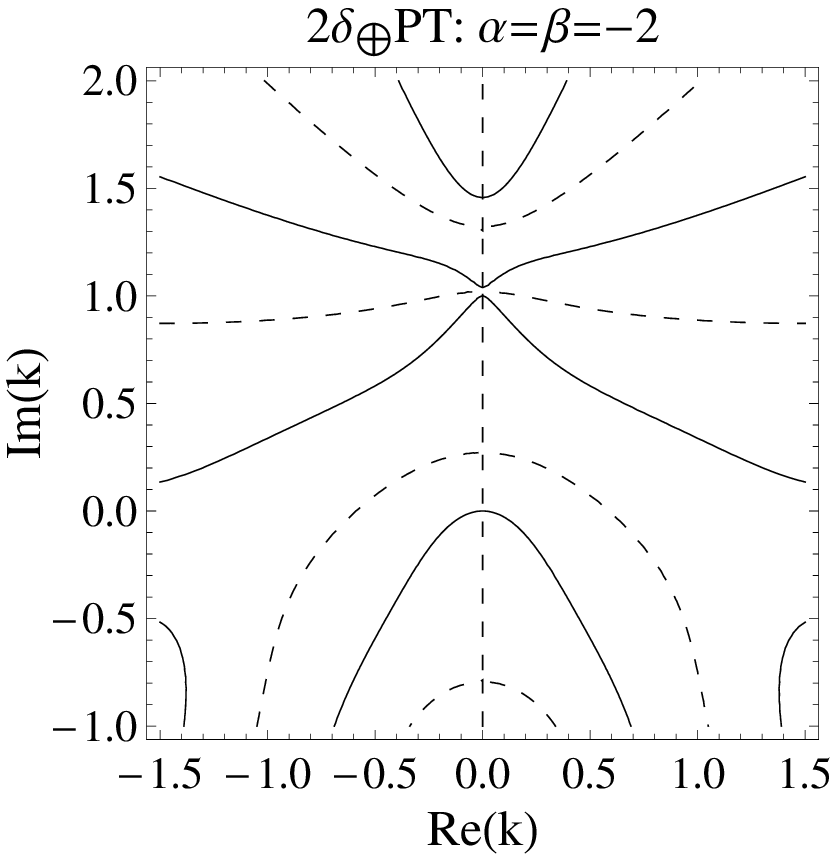}}
\caption{Plot of the curves: ${\rm Re}\Delta(k)=0$ (continuous lines) and
${\rm Im}\Delta(k)=0$ (dashed lines) (left). Plot of the curves: ${\rm Re}\Delta^{^{(K)}}(k)=0$ (continuous lines) and ${\rm Im}\Delta^{^{(K)}}(k)=0$ (dashed lines) (right). Strongly attractive case.}
\label{aa2}
\end{figure}

\begin{figure}[h]
  \centerline{\includegraphics[height=6cm]{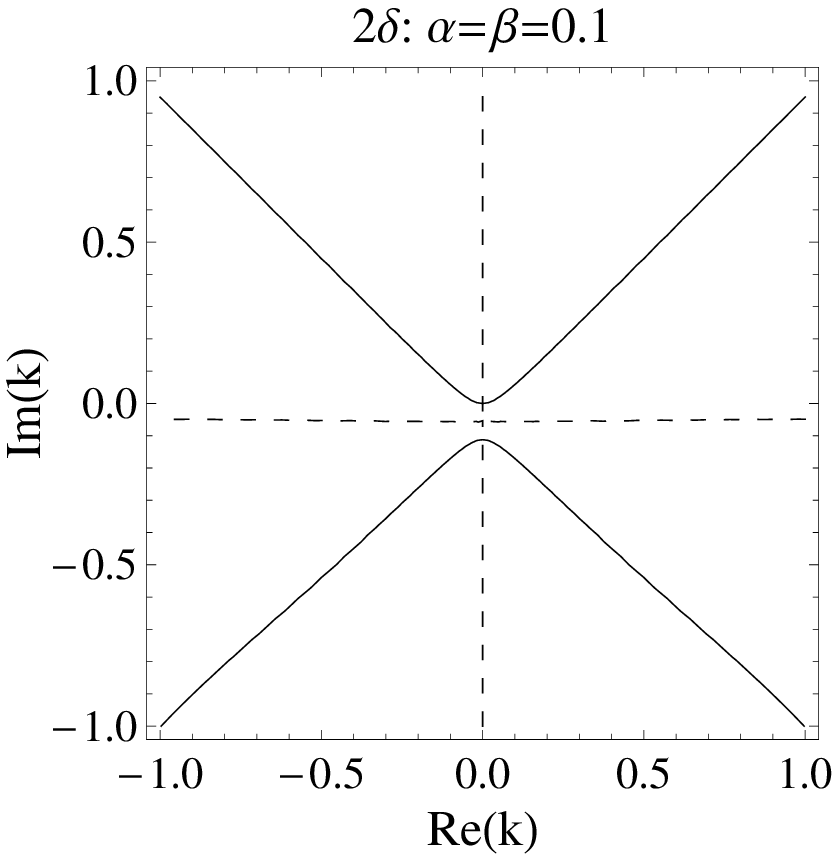}\quad\includegraphics[height=6cm]{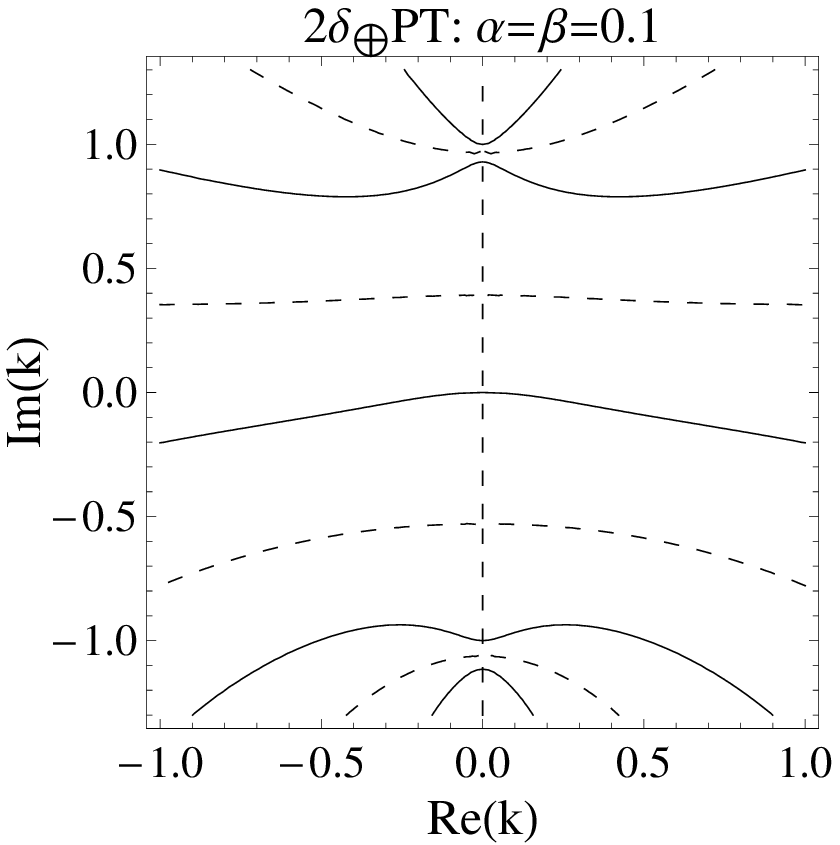}}
\caption{Plot of the curves: ${\rm Re}\Delta(k)=0$ (continuous lines) and
${\rm Im}\Delta(k)=0$ (dashed lines) (left). Plot of the curves: ${\rm Re}\Delta^{^{(K)}}(k)=0$ (continuous lines) and ${\rm Im}\Delta^{^{(K)}}(k)=0$ (dashed lines) (right). Weakly repulsive case.}
\label{rr01}
\end{figure}
In Figure \ref{rr01} the weakly repulsive case, $\alpha=\beta=0.1$, is depicted. There are no bound states in the two-$\delta$ walls configuration but both the real and imaginary part of $\Delta(k)$ are zero at one point in the negative imaginary axis due to the existence of one anti-bound state. The addition, however, of a PT well allows two bound states, zeroes of both ${\rm Re}\Delta^{^{(K)}}(k)$ and ${\rm Im}\Delta^{^{(K)}}(k)$ in the positive imaginary axis.

\begin{figure}[h]
  \centerline{\includegraphics[height=6cm]{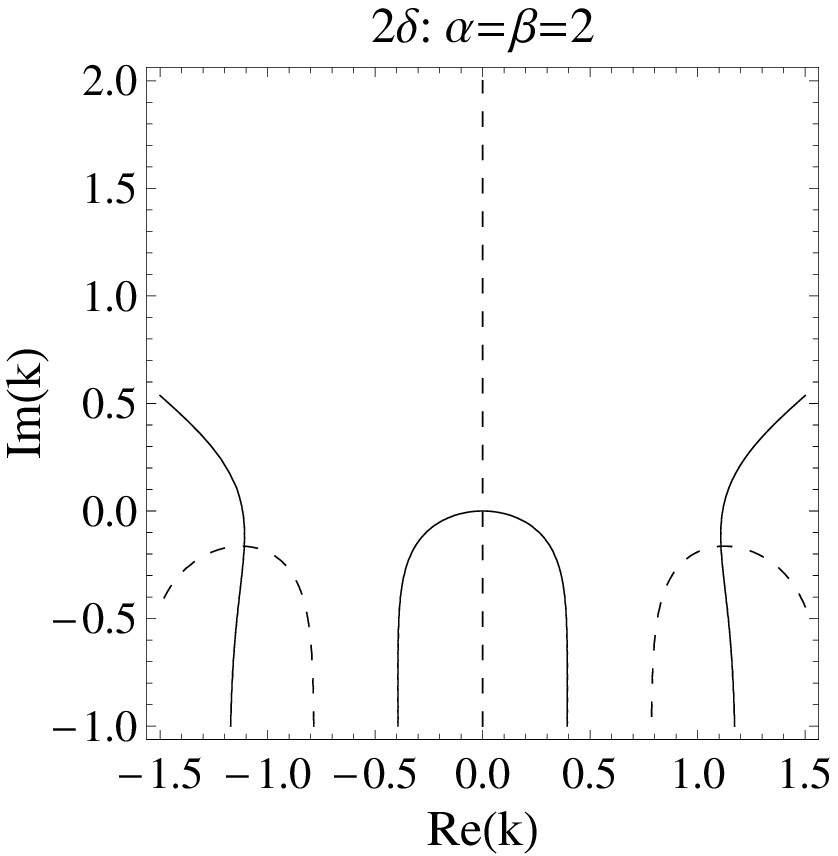}\quad\includegraphics[height=6cm]{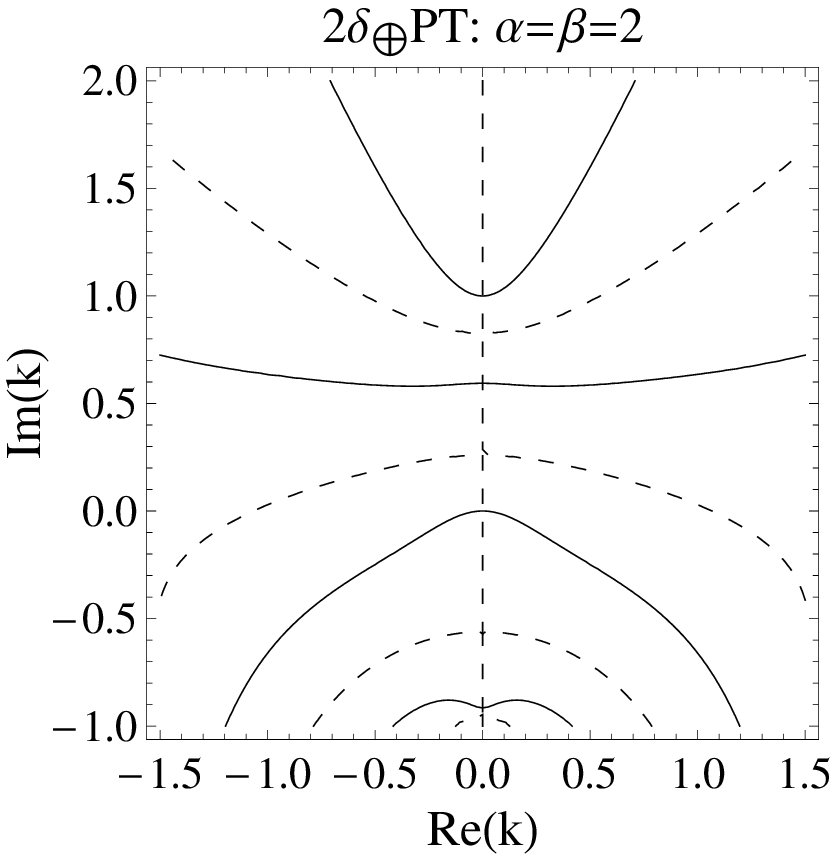}}
\caption{Plot of the curves: ${\rm Re}\Delta(k)=0$ (continuous lines) and
${\rm Im}\Delta(k)=0$ (dashed lines) (left). Plot of the curves: ${\rm Re}\Delta^{^{(K)}}(k)=0$ (continuous lines) and ${\rm Im}\Delta^{^{(K)}}(k)=0$ (dashed lines) (right). Strongly repulsive case.}
\label{rr2}
\end{figure}
Finally, we show in Figure \ref{rr2} the strongly repulsive case $\alpha=\beta=2$. There is a pair of resonances between the $\delta$ walls but two bound states arise when the PT well is included.

\section{Conclusions and outlook}

In sum, we draw the following conclusions:

\begin{itemize}

\item The one-dimensional scattering produced by two Dirac-$\delta$ potentials has been studied even
for the non-equal strength case.

\item In the limit of infinite strength walls the Casimir effect with Dirichlet boundary conditions has
been reproduced.

\item It has been shown that the analysis of the sine-Gordon kink one-loop fluctuations with Dirichlet
boundary conditions requires in this framework to consider two-$\delta$ plus a P$\ddot{\rm o}$sch-Teller
potential.

\item When the two $\delta$ walls are impenetrable the kink Casimir energy can be analyzed like the ideal
Casimir effect.

\end{itemize}

After completing this work we plan to include in the game also $\delta^\prime$ potentials in the hope
of finding more general boundary conditions.

%\begin{figure}
%  \includegraphics{example.eps}
%\end{figure}
%\begin{figure*}
%  \includegraphics[width=0.75\textwidth]{example.eps}
%\end{figure*}
%\begin{table}
%\begin{tabular}{lll}
%\hline\noalign{\smallskip}
%\end{tabular}
%\end{table}

%\begin{acknowledgements}
%\end{acknowledgements}



\end{document}